\documentstyle[psfig]{aa}
\def\ltsima{$\; \buildrel < \over \sim \;$}
\def\simlt{\lower.5ex\hbox{\ltsima}}
\def\gtsima{$\; \buildrel > \over \sim \;$}
\def\simgt{\lower.5ex\hbox{\gtsima}}
\def\gsimeq
{\hbox{\raise0.5ex\hbox{$>\lower1.06ex\hbox{$\kern-1.07em{\sim}$}$}}}
\def\lsimeq
{\hbox{\raise0.5ex\hbox{$<\lower1.06ex\hbox{$\kern-1.07em{\sim}$}$}}}

  \title{Mapping the inner regions of MCG-6-30-15 \\ with XMM-Newton}
    \author{G. Ponti
     \and      M. Cappi
     \and      M. Dadina 
     \and      G. Malaguti 
}

  \institute {{IASF-CNR, Sezione di Bologna, Via Gobetti 101, I-40129
               Bologna, Italy}}
   \offprints{Gabriele Ponti (ponti@bo.iasf.cnr.it)}
   \date{Received / Accepted }

\abstract{Timing analysis of a $\sim$95 ks long $XMM-Newton$ observation of the Seyfert 1 
galaxy MCG-6-30-15 is presented. Model-independent tools have been used with the intent of 
resolving the different components that produce the observed flux and spectral variations 
down to timescales as short as $\sim$300 s.
We find that the fractional variability is possibly due to a variable power law 
component in the hard band, and a slower thermal component at softer energies, 
but the data do not rule out effects due to the warm absorber.
The most relevant result of this work is the first detection of a significant 
enhancement of the fractional variability in the $\sim$4.7-5.8 keV energy band compared to the 
underlying continuum.
This represents one of the strongest, model-independent, pieces of evidence to date 
for the presence of redshifted relativistic matter accreting into the black hole.
During the brightest flare of the observation, 
a soft-to-hard time lag of $\sim$600 s is measured. A very significant increase 
of the iron line intensity is observed $\sim$3000 s after this flare.
In the framework of a disk-corona model and assuming the soft-to-hard time lag is 
due to Comptonization, these findings make it possible to estimate the electron density and the dimensions 
of the flare region and the disk-flare distance.

 }

\begin{document}

\maketitle

\section{Introduction}

According to the commonly accepted paradigm, Seyfert galaxies are thought to harbour at their 
center a supermassive black-hole surrounded by a geometrically thin accretion disk.
The hard power-law component that generally dominates the spectral emission above 2 keV 
is believed to arise in a hot plasma corona located above the surface of the accretion disk, 
where optical/UV photons from the disk are comptonized to X-ray energies. 
These X-rays in turn illuminate the disk, being either ``reflected'' towards the observer or thermalized back
into optical/UV emission (i.e. the so-called two-phase model, Haardt \& Maraschi 1991; 
Haardt \& Maraschi 1993; Haardt et al. 1997). 

Spectroscopic evidence for the two-phase disk-corona scenario is
seen in X-rays in the forms of a broad, fluorescent FeK line at 6.4 keV, 
an FeK edge at $\gsimeq$7 keV, a Compton "hump" at $\gsimeq$10 keV and a high-energy cutoff 
at E $\lsimeq$200 keV (e.g. Nandra \& Pounds 1994; Haardt 1997; Fabian et al. 2000; Perola et al. 2002). 
The unprecedented large collecting area of $XMM-Newton$ (Jansen et al. 2001), 
now offers for the first time the 
possibility of performing detailed, spectrally resolved, timing analysis of 
the brightest Seyfert galaxies. These studies potentially offer a method for 
constraining the emission models and their geometry 
(e.g. Edelson et al. 2002, Poutanen 2001).

MCG-6-30-15 (z=0.00775) is one of the brightest known Seyfert 1 galaxies.
It shows a very broad FeK line and has remarkable flux and spectral variability properties, 
with doubling times as short as $\sim$1000 s.
It has been observed twice by $XMM-Newton$: once in June 2000 for $\sim$100 ks 
(Wilms et al. 2001), and then in July 2001 for $\sim$300 ks (Fabian et al. 2002). 
In both observations, the time-averaged spectral analysis indicated the presence 
of a strong and broad FeK line extending down to about 3-4 keV, 
later detected also by Chandra (Lee et al. 2002).
Detailed model-independent temporal analysis of the longer 2001 
observation by Fabian et al. (2002) and Vaughan et al. (2003a) illustrated 
well the potential of this type of study.
The RMS spectrum presented by Fabian et al. (2002) showed a decrease 
in the variability at high energies interpreted as due to an interplay 
between a variable power law and a stable reflected component.
Vaughan et al. (2003a) showed that the power spectral density (PSD) of MCG-6-30-15 
is very similar to that of Galactic Black Hole Candidates (GBHCs), and detected 
a soft to hard time lag of about 200 s.
%In particular, they show the source broad band spectral variability 
%and the similarity of its PSD to the galactic black hole candidate ones. 
%They also found a soft-to-hard time lag of $\sim$ 200 s.

Here we present the temporal analysis of the June 2000 observation, with the data taken from 
the $XMM-Newton$ public archive. The PSD and structure functions 
are given in La Palombara et al. (2002). Those authors show evidence 
for the absence of any periodicity, and the presence of variability down to timescales of $\sim$300 s.
We perform here further analysis, with particular attention to the study of spectral 
variations and time lags during the whole observation and during a prominent and strong flare. 
The paper is organized as follows. Section 2 describes the data reduction.
In Sect. 3 the results of the analysis performed over the whole observation are presented,
while Sect. 4 presents the results during the strongest flare.
Thermal Comptonization models predictions are compared with obtained results in Sect. 5, 
while in Sect. 6 we attempt to give constraints on some physical parameters (e.g. 
density and dimensions of the disk-corona system). Alternative models are briefly described in Sect. 7, 
followed by conclusions in Sect. 8.

\section{Observations and Data Reduction}

The data were collected during the observation performed on June 11$^{th}$ 2000, with the EPIC MOS1, MOS2 and pn cameras (Str\"uder et al. 2001) operated  
in timing, full-frame, and small window modes, respectively. 
MOS1 timing data were excluded from the analysis because it was strongly contaminated 
by the high background.
MOS2 data were also excluded because they suffered from substantial pile-up. 
In the following we thus used only the pn data.

The data were reduced and screened using the SAS software v5.3. 
The analysis of the pn data was performed using both single and double events (i.e. pattern $\leq$4).
%in a standard manner using all 
%pn events with pattern $\leq$4 {\bfseries that correspond to both single and double events}.
An {\it a-posteriori} check confirmed that the results are identical using only single events.
The last $\sim$5 ks of the observation were excluded from the analysis because 
strongly contaminated by soft-{\it p}$^{+}$ flares.
Source counts were extracted from a circular region of 
45$\arcsec$ radius centered on MCG-6-30-15, while the background events were taken 
from two larger rectangular regions far from the source.
The total useful exposure time was $\sim$95 ks, yielding a total number of 
$\sim$$1.2\times 10^{6}$ source counts in the 0.2-10 keV band.
The average 2-10 keV flux of $F_{2-10}$ = 2.4$\times 10^{-11}$ erg cm$^{-2}$ s$^{-1}$ is somewhat 
low for MCG-6-30-15, being only 20\% higher than during the ``deep minimum state'' found by 
Iwasawa et al. (1996). Furthermore, during the observation, the 2-10 keV flux varied in the range 
1.5-4.8$\times 10^{-11}$ erg cm$^{-2}$ s$^{-1}$.

\section{Analysis of the whole observation}

Figure 1 shows the 0.2-2 keV (upper panel), 2-10 keV (middle panel) light 
curves and their ratio (lower panel) as a function of time.
As already known, the source is highly variable with variations up to a 
factor of $\sim$4.
The light curves show many shots/bursts, with a large prominent 
flare that occurred just before the end of the observation at {\it t} $\sim$9.2$\times$10$^{4}$ s 
(see Figure 1, upper panel). 
Hereinafter we will refer to this event as the ``Flare''. 

Figure 1 (lower panel) also clearly shows that this flux variability is associated to 
strong spectral changes, confirming previous findings on Seyfert galaxies 
(e.g., Mushotzky et al. 1993; Lee et al. 2000; Shih et al. 2002) 
that the spectrum becomes softer as the flux increases.
%{\bfseries in particular it shows that the spectrum softens with flux}, 
%confirming previous findings {\bfseries seen in this and other Seyfert galaxies (e.g., Mushotzky et al. 1993;} 
%Lee et al. 2000; Shih, Iwasawa and Fabian 2002).  
 
\begin{figure} [t]
\psfig{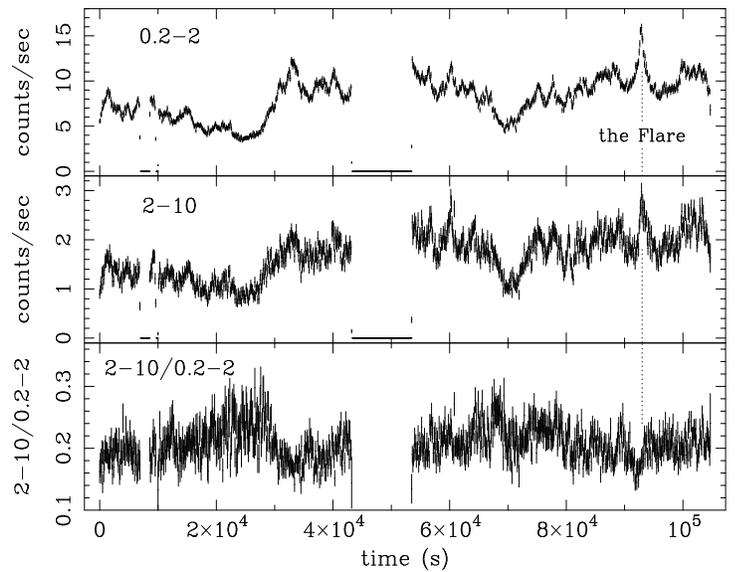}
\caption{($Upper~ panel$) Soft (0.2-2 keV) and ($middle~ panel$) hard (2-10 keV) EPIC pn light curves with 100 s time bins. 
($Lower~ panel$) Hardness ratios ($\frac{2-10 keV}{0.2-2 keV}$) as a function of time. Note that count rates are not corrected 
for the ``live time'' ($\sim$71\%) for the pn small window operating mode (Str\"uder et al. 2001).}
\end{figure}

\subsection{Fractional Variability or ``RMS Spectrum''}

To characterize the source variability as a function of energy, we used the 
root mean square variability function, also called ``RMS spectrum'' (for definition and details, 
see the Appendix and Edelson et al. 2002). Similar to the more commonly used 
``excess variance'' (Nandra et al. 1997, Edelson et al. 2001), the RMS spectrum allows the quantification 
of the source fractional variability as a function of energy in a model-independent way.
The advantage of the RMS spectrum (as defined in the Appendix) is that its ``modified'' treatment of the 
errors allows a finer energy binning, thus yielding higher sensitivity to fine spectral structures.

Figure 2 shows the obtained normalized RMS spectrum. 
It clearly shows three different gross ``patterns'': a soft, smooth increase of the 
fractional variability from about 0.2 to 0.6 keV; a ``plateau'' of maximum variability 
between $\sim$0.6--2 keV; a gradual decrease of the variability 
at E $\gsimeq$2 keV (of the form $\propto$$E^{-0.16}$). 
Overall, the gross shape and normalization of this RMS spectrum is consistent 
with that observed one year later during the 2001 observation (Fabian et al. 2002). 
In particular, the normalization of the RMS spectrum is slightly less than the one 
from Fabian et al. (2002). 
This is in agreement with the flux-rms correlation found in Vaughan et al (2003a).

Most importantly, thanks to the finer sampling achieved at high energies, we detect for the first time 
some fine structures above 5 keV, i.e. close to the FeK complex: a steep rise of 
variability in the $\sim$4.7-5.8 keV energy band followed by a drop at energies close to 6.4 keV, 
and a flattening at energies greater than $\sim$7 keV.
To estimate the significance of the variability rise in the 4.7-5.8 keV energy band, 
we first fitted the 2-10 keV RMS spectrum with a power-law continuum. 
The fit then improved significantly ($\Delta \chi^2 \sim$ 15, i.e. $>$99\% confidence in a F-test) 
when adding to the model a gaussian line at 5.3 keV, confirming that this excess is highly significant.  
A detailed discussion and interpretation of the source variability is given in Sect. 5.1.

\begin{figure} [t]
\psfig{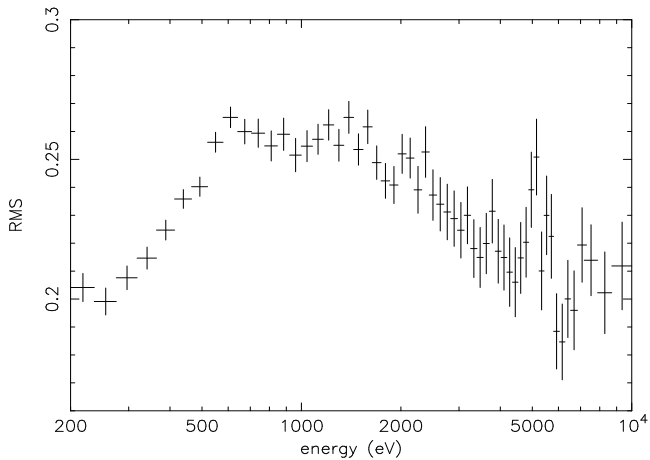}
\caption{RMS spectrum calculated with time bins of 6000 s and with energy bins grouped in order 
to have $>$350 counts per bin. Our choice undersamples the detector energy resolution at low energies. 
Errors are 1$\sigma$ (see Appendix).}
\end{figure}

\subsection{Study of the Time Lags}

To search for time lags, we used the cross correlation 
function (CCF; White \& Peterson 1994) 
that we calculated using the $crosscor$ package in 
FTOOLS v5.2 (this command implements the DCF, e.g. Edelson \& Krolik 1988, 
in the ``correl'' routine of the ``Numerical Recipes'', Press et al. 1992).  

The energy intervals for the CCF were chosen so as to sample the RMS variability pattern. 
Light curves were, thus, extracted for the following energy bands:  
E$_1$ (0.2-0.6 keV), E$_2$ (0.6-2.2 keV), E$_3$ (2.2-2.6 keV), E$_4$ (2.6-4.5 keV), 
E$_5$ (4.5-6.8 keV), and E$_6$ (6.8-10 keV). 

Due to the presence of an instrumental switch-off in the middle of the observation 
(see Fig. 1), the CCF study has been performed separately, before and after it.
Moreover, we calculated the CCFs with and without the Flare period. 
E$_2$ was used as the energy band of reference for all CCFs because the light curve in this band 
had highest statistics.

Figure 3 shows, as example, the CCFs obtained during the 33 ks after the switch-off
(which does not include the Flare period). The CCFs are fairly wide and 
show no significant time lag between the different energy bands, with an upper limit of 
about 200 s. The CCF appear also to be somewhat asymmetric, with a marginal excess of 
correlation moving from negative to positive time lags as the energy bands go from soft to 
hard (i.e. from top to down in Fig. 3). Indicative could also be the 
possible secondary peak in the hardest band corresponding to a time delay of $\sim$1.5-2 ks.
Similar results have been found, extended to higher energies, 
by the analysis of the simultaneous XMM-Newton and RXTE data 
(J. Wilms, private communication). 
However, in our data the skewness parameter does not give any significant 
deviation from a symmetric profile. 
Better statistics are thus needed for any firmer conclusion on this point.  
%Similar (inconclusive) results were obtained for 
CCF analysis of only the first part of the observation give the same inconclusive results.
%, as 
\begin{figure} [t]
\psfig{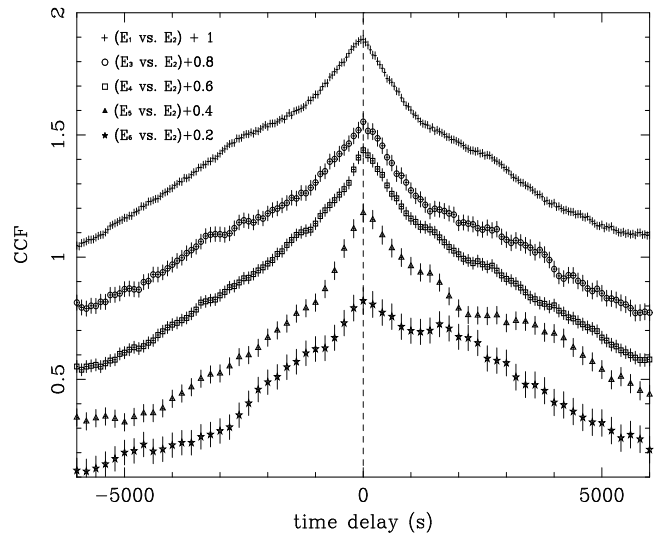} 
\caption{Rescaled CCFs calculated in different energy bands (E$_1$:0.2-0.6 keV, E$_2$:0.6-2.2 keV, 
E$_3$: 2.2-2.6 keV, E$_4$:2.6-4.5 keV, E$_5$:4.5-6.8 keV and E$_6$:6.8-10 keV) during the 33 ks 
following the instrumental switch-off. Values of CCFs have been increased by 1, 0.8, 0.6, 0.4 and 0.2 
going from soft to hard (top to down) for clarity purposes.}
\end{figure}
%well as on other shorter periods (e.g. the one between t$\sim$2$\times$10$^{4}$--3.5$\times$10$^{4}$s), 
%except for the Flare period.

\section{Analysis of the Flare period}

The present observation is characterized by the presence of 
a bright flare, an enlargement of which is shown in Fig. 4. 
In studying this event one hopes to isolate the physical mechanism(s) responsible for the variability 
so as to constrain emission models and, possibly, 
generalize these findings to other 
periods when the flares are more elusive.

\subsection{Spectral variability during the Flare}

The light curve hardness ratios shown in Fig. 1 (Sect. 3) clearly demonstrate that 
MCG-6-30-15 exhibits spectral variations during the Flare. 
To quantify this effect, we first produced a set of ratios (commonly called ``PHA ratios'') 
\begin{figure} [t]
\psfig{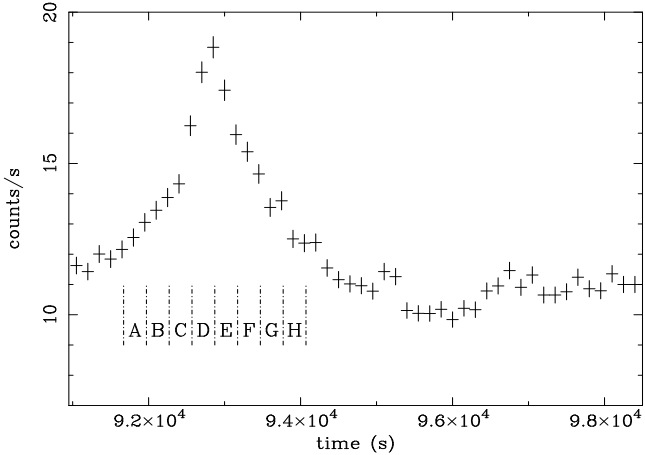}
\caption{Enlargement of the Flare period (see also Fig. 1). The light curve is 
computed between 0.2-10 keV with time bins of 150 s. Periods during the Flare where 
PHA ratios have been calculated (see Sect. 4.1 and Fig. 5) are labelled from A to H, and are 300s long.} 
\end{figure}
between spectra acquired in 8 subsequent time intervals of 300 s during the 
Flare (see Figures 4 and 5) and the spectrum accumulated during the whole observation.
Figure 5 shows the results of such an analysis.
Before the Flare (panel A), the source shows a spectrum close to the average one. 
At the beginning of the Flare, the source shows an increase of the soft counts (panel B, C and D) 
followed by a delayed response at higher energies (panel E and F). Finally, the source gradually 
returns to its initial spectral state (panel G and H). Overall, this result suggests the presence of a 
soft-to-hard time lag. 

To quantify this effect in more details, we performed a CCF analysis limited to the Flare period 
using the same energy bands defined in Sect. 3.2.
As shown in Fig. 6 (left panel), we find that the harder light curves systematically lag 
the softer one by $\sim$ 80 s, 260 s, 310 s, 340 s, and 480 s, respectively.
This soft-to-hard lag is even clearer from Fig. 6 (right panel)
where the peak time lag is reported as a function of energy\footnote{
To calculate the errors associated to the time lag, we 
followed two different methods: CCF peaks were first ``$\chi^2$''-fitted 
with a Gaussian function and the errors 
were then estimated by assuming i) a $\Delta \chi^2$= 4, and ii)  
$\frac{\sigma}{\sqrt{N}}$, where $\sigma$ is the FWHM of the 
Gaussian and N is the number of time bins considered in the fit. 
Most conservative values obtained following either method 
were used and reported in Fig. 6.}. The time lag clearly increases with energy 
($\tau_{{\rm lag}} \sim 0.071 \times E$; where E is in eV and the time lag is in s).

\begin{figure*} [t]
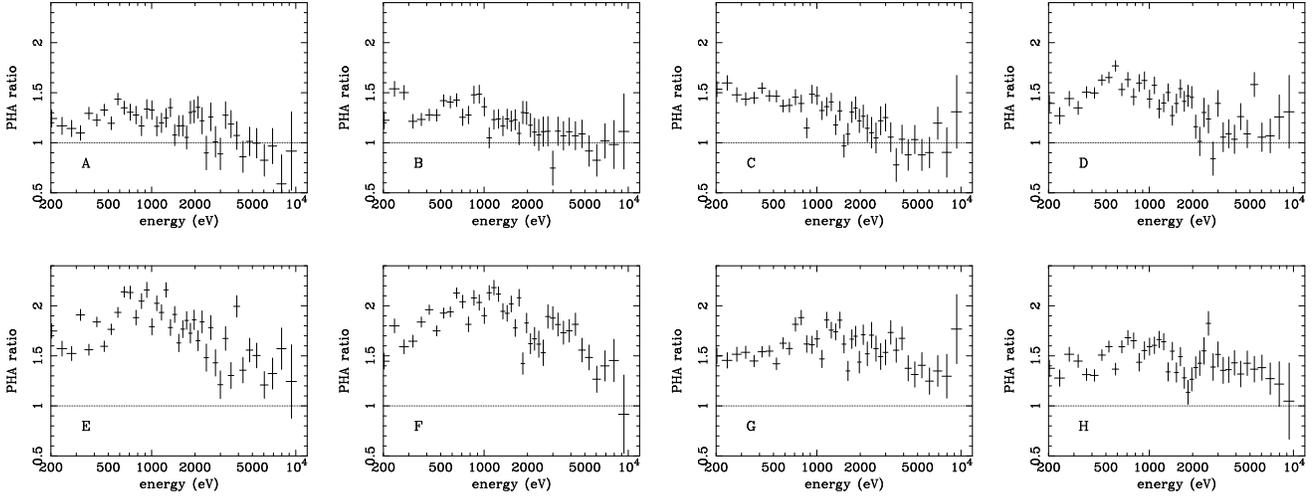

\begin{tabular}{c c c c}
\psfig{file=fig_mcg6de1.ps,width=4cm,height=3cm,angle=-90} & \psfig{file=fig_mcg6de2.ps,width=4cm,height=3cm,angle=-90} & \psfig{file=fig_mcg6de3.ps,width=4cm,height=3cm,angle=-90} & \psfig{file=fig_mcg6de4.ps,width=4cm,height=3cm,angle=-90} \\
\huge
&&&\\
\small
\psfig{file=fig_mcg6de5.ps,width=4cm,height=3cm,angle=-90} & \psfig{file=fig_mcg6de6.ps,width=4cm,height=3cm,angle=-90} & \psfig{file=fig_mcg6de7.ps,width=4cm,height=3cm,angle=-90} & \psfig{file=fig_mcg6de8.ps,width=4cm,height=3cm,angle=-90} \\
\end{tabular}
\caption{MCG-6-30-15 spectral variations during the Flare. 
PHA ratios are calculated by dividing spectra acquired in subsequent time intervals of 300 s of about 4500 counts (see Fig. 4) 
and the average spectrum of the entire observation. This figure illustrates the gradual softening, followed by a hardening 
and subsequent decrease of the spectrum during the Flare. }
\end{figure*}

\begin{figure*} [ht]
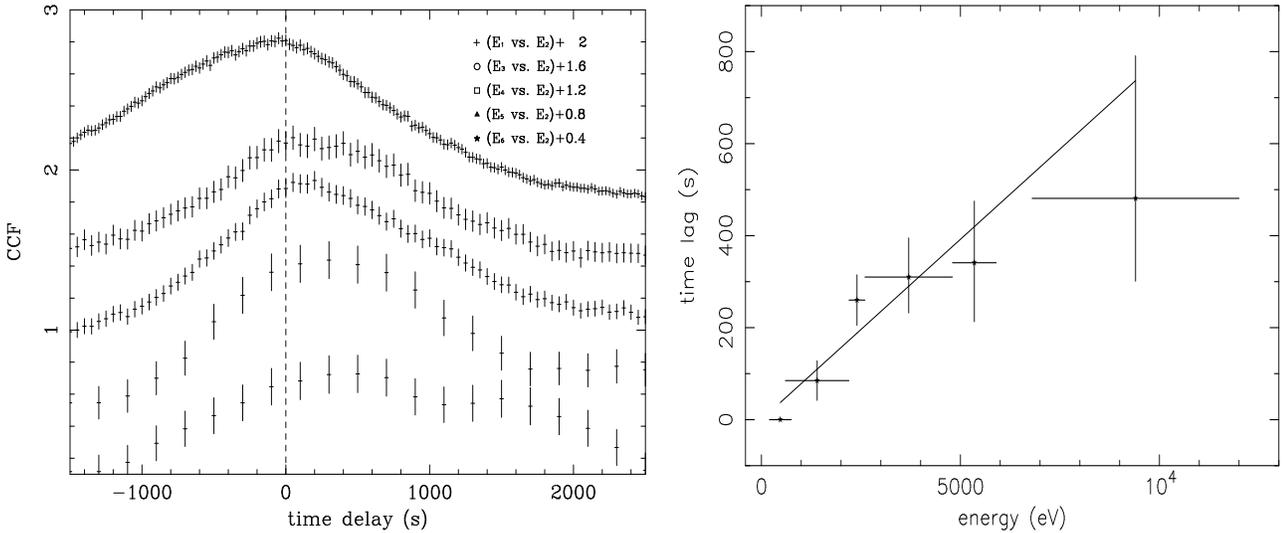

\begin{tabular} {c c}
\psfig{file=CCFflareretta.ps,width=8.5cm,height=7cm,angle=-90} & \psfig{file=time_lag.ps,width=8.0cm,height=7cm,angle=-90}  \\
\end{tabular}
\caption{($Left~ panel$) CCFs between light curves extracted during the Flare. Energy bands are as in 
Fig. 3, and values have been increased by 2, 1.6, 1.2, 0.8 and 0.4 going from soft/top to hard/down for clarity purposes.
The dashed vertical line indicates a zero lag value. 
For each energy band, the differences between the peaks of CCF distributions 
and the dashed vertical line indicate the time lags. 
($Right~ panel$) Time lag (in s) vs. energy (in eV) is shown. 
Peak time lag values and associated (68\% confidence) errors have been estimated 
through a $\chi^2$-fitting method (see footnote n. 1).}
\end{figure*}

This result is consistent with the time lag ($\sim$ 200 s) that was found during the 2001 long 
($\sim$300 ks) $XMM-Newton$ observation by Vaughan et al. (2003a). 
In that case, however, the time lag was found only on timescales longer than $\sim$10$^{4}$ s. 
This was interpreted as evidence for a drop of coherence on shorter timescales. 
The fact that we detect it in a single Flare event (during which 
the source is likely to be dominated by a single active region, thus with maximum coherence), 
coupled with the null result on the whole observation (Sect. 3.2) confirms and strengthens this 
hypothesis.

%found during the 2001 $XMM-Newton$ observation, but 
%on timescales longer than $\sim$10$^{4}$s (Vaughan et al. 2003). But, unlike Vaughan et al., no time lag 
%was found here when considering the whole observation. This could however be explained by a drop of the source 
%coherence on shorter time-scales (whenever the emission is not dominated by a single flare) 
%that would hide such a lag.
%the corona has no longer a simple transfer function, because it has much active regions with temporally and spatially different physical properties.

\subsection{FeK reverberation in response to the Flare?}

To study the FeK line variations during the Flare period 
in a model-independent way, we extracted the source spectrum in three time intervals (1000 s long) 
before, during and after the Flare (we will refer to these time intervals 
as the PreFlare, Flare and PostFlare periods).
The PHA ratios between the PostFlare/PreFlare periods highlights a strong line 
increase that took place during the PostFlare period.
During the Flare only a softening of the continuum is visible.
To further investigate the variations, we performed a spectral analysis 
on a set of eight subsequent time intervals of $\sim$1000 s each around the Flare.
\begin{figure*} [ht]
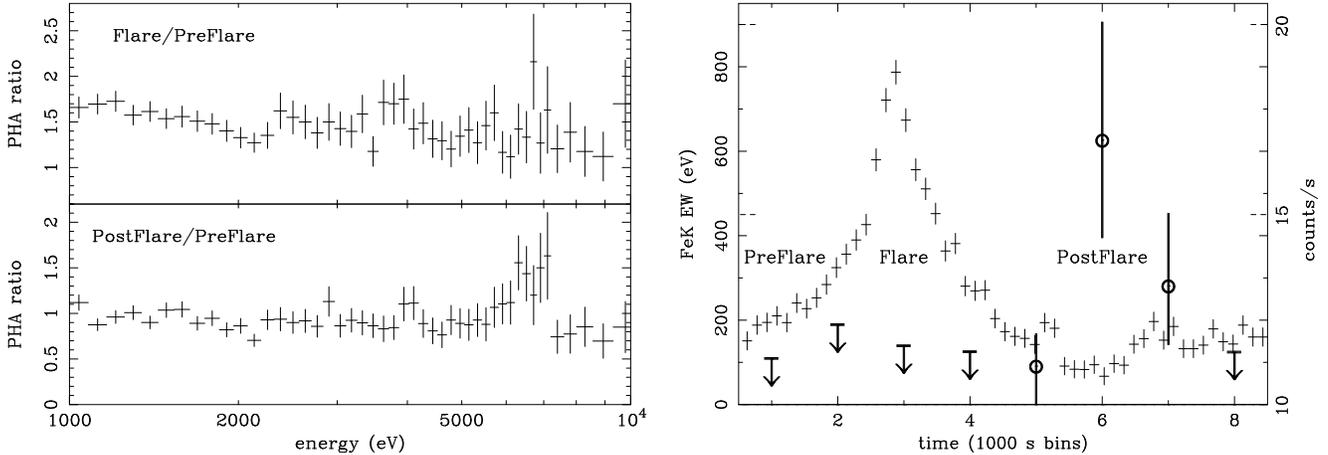

\begin{tabular}{c c}
\psfig{file=ratio.ps,width=8.5cm,height=6cm,angle=-90} & \psfig{file=fig_mcg8tlsum.ps,width=8.5cm,height=6cm,angle=-90} \\
\end{tabular}
\caption{($Left~panel$) Flare/PreFlare and PostFlare/PreFlare PHA 
ratios. ($Right~panel$) FeK$\alpha$ line equivalent width (in eV, left y axis) 
calculated in 1000 s spectra, over-plotted on light curve (right y axis) from Fig. 4. 
Errors are at 90\% confidence level for one interesting parameter.}
\end{figure*}
We fitted these data using 
%we computed the ratios of 
%the two spectra (1000 s long) obtained during and before the Flare (Flare/PreFlare) 
%and the ratio between after and before the Flare spectra (PostFlare/PreFlare).}
%Figure 7 (Left Panel) clearly indicates that 
%the iron line flux increased significantly $\sim$3000 s after the Flare, {\bfseries in fact,
%the PostFlare/PreFlare PHA ratio shows a strong variation of the line while the continuum 
%is similar. On the other hand, no line variations are evident before and during the Flare, 
%we could see only that the spectrum is softer during the Flare. 
%To obtain further (but model-dependent) informations on any possible variation of the 
%FeK line complex, we accumulated spectra in successive 1000 s bins around the Flare, 
%so as to have at least $\sim$10000 counts for each time interval.
%Then we fitted the 2-10 keV data with 
a simple power-law model 
($\Gamma \sim$1.75-1.95) plus a Gaussian line at $\sim$ 6.4 keV.
Figure 7 (Right Panel) clearly shows that only upper-limits on the EW 
(of $\sim$100-200 eV) were obtained for the FeK line before and during the Flare.
On the other hand, the line recorded after the Flare (periods 6 and 7 in Fig. 7) 
was remarkably strong (EW$\sim$680$\pm$200 eV and 280$\pm$150 eV) and 
broad ($\sigma$ $\sim$0.39$\pm$0.20 and 0.19$\pm$0.12 keV, respectively).
It should be noted that a similar effect was found for the absolute flux 
of the FeK line, with $A_{{\rm FeK}} < 6.5 \times 10^{-5}$; $=1.8^{+1}_{-0.3} \times 10^{-4}$; 
$ < 1.6 \times 10^{-5}$ photons cm$^{-2}$ s$^{-2}$ 
at periods 5, 6 and 7 in Fig. 7(right), respectively.
More complex models for the line (e.g. a DISKLINE model; Fabian et al 1989) 
have not been applied given the low signal-to-noise ratio of these 1000 s spectra.

It is interesting to note that such spectral changes, detected here, 
are qualitatively similar to those found by ASCA in 1994 during another remarkably bright flare 
(Negoro et al. 1999; Negoro et al. 2000)

\section{Implications of the observed spectral variability}

The X-ray emission mechanism operating in Seyfert galaxies is usually thought to be  
inverse Compton on seed soft photons that are iteratively up-scattered by a hot electron corona 
(e.g. Sunyaev \& Titarchuk 1980; Haardt \& Maraschi 1991; 
Haardt, Maraschi \& Ghisellini 1997). 
We test this hypothesis against the spectral variations measured during the 
$XMM-Newton$ observation.

\subsection{Constraints from the long ($\gsimeq$ 5 ks) timescale variability: Continuum and FeK line}

%As previously stated, the continuum variations seen in the RMS spectrum (Figure 2 in $\S$3.1) are grossly consistent with those 
%obtained during the 2001 observation (Fabian et al. 2002). 
%
As noted earlier (Sect. 3.1), the RMS spectrum allows, for the first time, 
the detection of fine structures at the energy of the FeK line complex.
In particular, the drop in variability at $\sim$6.4 keV is consistent with a fraction of the 
FeK$\alpha$ line being produced in outer (less variable) parts of the accretion disk 
(see also Fabian et al. 2002; Lee et al. 2002).
The variability excess located between 4.7 and 5.8 keV instead implies the presence of an
intense, broad spectral component that varies more than the underlying continuum.
This excess could be due to a broad, 
redshifted iron line component produced by fluorescence from an accretion disk around either 
a Schwarzschild or a Kerr black hole. 
If so, the emitting iron should have a bulk velocity of about 0.3-0.4 c. 
Another, maybe more extreme, possibility could be that the excess of variability is 
associated to absorption, rather than emission, from almost fully ionized iron falling into 
the black hole (along our line of sight) at about $\sim$0.2-0.4 c 
(see e.g. Nandra et al. 1999, Ruszkowski \& Fabian 2000, Longinotti et al. 2003).
Whatever its precise origin, the excess of fractional variability detected here 
between $\sim$4.7 and 5.8 keV represents strong (because model-independent) 
evidence for the presence of relativistic matter accreting (because redshifted) 
onto a black hole.

In the soft band, the sharp drop of variability at E $\lsimeq$0.6 keV is likely 
due to the rise of a sharp soft-excess component clearly seen 
in the source-average spectrum (e.g. Fabian et al. 2002; Pounds \& Reeves 2002). 
This feature, best fitted by a black-body component of average temperature 
kT $\sim$100 eV, is consistent with either intrinsic or reprocessed 
thermal emission from the less variable optically thick accretion disk. 
On the reprocessed emission hypothesis, the data do not exclude 
the presence of soft X-ray disklines produced from the inner, more variable, 
disk regions which may contribute between $\sim$ 0.6-1.0 keV 
(Branduardi-Raymont et al. 2001; Lee et al. 2001; Sako et al. 2003; Turner et al. 2003).
Another contribution to the variability in this band could come from the warm absorber that has been 
observed to vary on timescales down to 10$^{4}$ s (Otani et al. 1996), and therefore 
should be observable by the RMS spectrum.

\vspace{-0.15cm}
In the hard band, the drop of variability could be due to different interplaying phenomena. 
Phenomenologically, it could be explained as a single power-law spectrum that varies 
with a pivot point at E $\gsimeq$10 keV, as the analysis of the hardness ratio indicates  
(the spectral index softens as the flux increases).
Physically, such behavior is consistent with what is expected if the coronal optical 
depth varies while the dissipated luminosity remains constant, and it is just 
what is expected for the disk corona models 
as long as the spectral index $\Gamma_{[2-10]}$ is  $\lsimeq$2 (Haardt et al. 1997).
Another possibility could be to have a two-component continuum model like the one 
proposed by Shih et al. (2002), Fabian \& Vaughan (2003) and 
Miniutti et al. (2003), which foresees one steep power law rigidly variable plus a flatter 
and almost constant reflection-dominated component. 
In this model, the reflection component is produced very near the black hole 
and is heavily affected by light bending.
As a result, the emission line can be strong down to $\sim$3-4 keV (Fabian et al. 2002), 
and could reduce the continuum variability in a complex pattern. 
Another possible explanation for the complex FeK line versus continuum variability is 
the ionized reflection model proposed by Ballantyne et al. (2003) 
which requires two distinct reflectors: one from the inner accretion disk to build up the red wing 
and the other from the outer disk to fit the blue peak.
Alternatively, some contribution to the variability decrease at hard energies could also come from 
the Comptonization process itself. In fact, higher energy photons, after a great number of interactions 
(i.e. at larger distances and in different physical states), may display a smoother 
variability than do softer photons.

\par\noindent

\subsection{Constraints from the short ($\sim$ 100 s) timescales variability: Continuum and FeK line}

The spectral variability during the strong ($\Delta$F$\sim$2 in $\sim$1000 s) Flare is characterized by 
a soft-to-hard time lag of a few hundred seconds (see Fig. 6, right panel). If the Flare time lag, 
$\tau_{{\rm lag}} \sim$600 s, is interpreted as the delay required by Comptonization to upscatter soft 
photons up to 10 keV (see also \S 6.1), then it is comparable with, or even larger than, the minimum 
time scale ($\sim$300 s) of the variability. 

The remarkably strong (EW $\sim$680 eV) FeK line is detected 
with a delay of $\sim$3000 s after the Flare (Fig. 7, right panel), 
which may be indicative of a disk reverberation 
in response to the Flare. This indicates that simulations should also incorporate this geometric 
delay when computing the expected spectral variations. 
Moreover, we note that, just after the Flare, the FeK complex 
is present up to $\sim$7.0 keV. This suggests that the Flare may have been responsible for 
ionizing the disk and/or illuminating its approaching (blue-shifted) parts.

\section{Mapping the physical parameters of the disk-corona region}

In the framework of a hot corona model scenario, and assuming that the lag observed during the Flare is due to 
Comptonization, it is possible to derive a number of physical and geometrical parameters 
of the disk-corona system, as follows.

\subsection{Dimensions and density of the Flare region}

Assuming that the electron temperature in the emission region, best constrained by the 
high-energy BeppoSAX data (Guainazzi et al. 1999; Fabian et al. 2002), is in the 
range $\sim$100-300 keV, seed soft 
photons emerging from the disk will need about 5-10 interactions\footnote{
We estimated the number of interactions that a photon undergoes to be up-scattered from $\sim$200 eV to 10 keV 
by iterating the equation $h\nu^{''}\simeq \gamma_e^2 h\nu$ that links the average photon energy before ($h\nu$) 
and after ($h\nu^{''}$) an inverse-Compton scattering event. In this equation, 
$\gamma_e$ is the electron Lorentz factor given by the equation: 
$E_{e}=m_ec^{2}(\gamma_e-1)$, and we assumed that the electron energy ($E_{e}$) 
is equal to the high-energy cut-off of the X-ray spectrum.}
($N_{{\rm int}}$) to be up-scattered from $\sim$200 eV to 10 keV. 
Thus, the mean free path of 10 keV photons ($\lambda_{10}$) 
is linked to the Comptonization timescale ($t_{{\rm Compt}}$) through the 
equation: 
$\lambda_{10}~=~t_{{\rm Compt}} \times ~c~=~\frac{\tau_{{\rm lag}} \times ~c}{N_{{\rm int}}}\sim2-4\times 10^{12} {\rm ~cm}$,
where $\tau_{{\rm lag}}$ is the time lag ($\sim$600 s) between the 0.2 and 10 keV emission 
and $c$ is the speed of light. Assuming a random walk motion and $N_{{\rm int}}$=10, 
this implies a displacement from their original production region of 
$\sim \sqrt{N_{{\rm int}}}~ \lambda_{10} \sim 6 \times 10^{12}$ cm. 
Therefore it is reasonable to assume that, when the bulk of 10 keV photons are produced, 
the Flare dimensions ($R_{{\rm Flare}}$) should be of the order of $\sim 6 \times 10^{12}$ cm.
For values of the optical depth ($\tau$) ranging between $\tau \sim 0.1-1$,
a typical range for Seyfert galaxies including MCG-6-30-15 (e.g. Petrucci \& Dadina 2003), 
this yields also a rough estimate of the Flare region electron density ($\rho_{e^-}$):
$\rho_{e^-}=\frac{\tau}{\sigma_{\rm T} R_{{\rm Flare}}} \sim 2.5-25\times 10^{10}{\rm ~cm}^{-3}$,
where $\sigma_T$ is the Thomson cross-section.

The dimensions of the Flare region ($R_{{\rm Flare}} \sim 
6 \times 10^{12}$ cm) estimated above correspond to $\sim$ 40 gravitational radii ($r_{\rm g}$) assuming 
the mass of the black hole in MCG-6-30-15 to be $M_{{\rm BH}}\sim10^6 M_{\odot}$ (e.g. Vaughan et al. 
2003a). This value is consistent with 
the limit implied by causality arguments for the factor 2 variation in $\sim$1000 s 
during the Flare (Fig. 4), which gives $R_{{\rm Flare}} \lsimeq ~c\times t \simeq 3\times 10^{13}$ cm 
$\simeq$ 200 $r_{\rm g}$.

\subsection{Measure of the disk-corona distance}

The spectral analysis performed during and after 
the Flare (Fig. 7) clearly demonstrates a significant increase of the 
FeK line intensity $\sim$3000 s after the Flare peak.
Assuming that the line is produced in the accretion disk and in response to the Flare, it is 
possible to estimate the distance between the Flare emission region and the disk. 
To perform this accurately, one should know when precisely the intensity of the line started to increase, 
the transfer function of the disk and the physical condition of the Flare region. These are all unknowns 
but, if ever significant, they would introduce delays. Thus the estimate given below shall be considered 
only an upper limit.
  
Since the hard radiation ionizing the iron on the disk surface lags 
the Flare peak by about $\sim$600 s ($\S4.1$, Fig. 6), the upper-limit on the distance between the Flare 
region and the disk ($\Delta l_{{\rm F-d}}$) is given by:
%\begin{displaymath}
$\Delta l_{{\rm F-d}}~\lsimeq~\frac{3000 -600}{2}\times c = 4\times 10^{13} {\rm cm} \sim 240~ r_{\rm g}$,
%\end{displaymath}
also consistent with the marginal presence of a second peak in the CCFs 
at higher energies discussed in $\S3.2$, Fig. 3.

\section{Other models for the observed spectral variability}

Instead of considering models with a single corona (as in the earliest two-phase models) one could also 
consider patchy corona models (e.g. Haardt et al. 1994), i.e. models in which 
several active regions are present, with (more or less) independent physical (i.e. spectral) evolution.
In this scenario, the short timescale variability is expected to be dominated by the evolution of 
the individual active regions. On long timescales, these models must be able to reproduce the 
observed broad-band spectra that clearly suggest thermal Comptonization.
However, the single flare event could be dominated by different, non-thermal, 
mechanisms. In fact, there are problems in maintaining a thermal plasma in rapidly varying sources, 
since it may not have time to thermalize (Ghisellini et al. 1993).

Alternative mechanisms could for example involve magnetic reconnection above the disk surface 
(Haardt et al. 1994; Stern et al. 1995; Poutanen \& Svensson 1996; 
Beloborodov 1999a,b; di Matteo et al. 1999). 
Such mechanisms can introduce changes in the energy dissipation rate and/or in the geometry 
of the flaring regions (e.g. their distance from the disk) and could therefore 
reproduce the soft-to-hard spectral evolution found in the present paper (e.g. Poutanen \& Fabian 1999).
Electron-positron pairs could also contribute significantly to the coronal optical depth during the 
strongest flare periods. Indeed, using the standard equation for the compactness 
$l=\frac{L}{R}\frac{\sigma_{T}}{m_e c^3}$ (assuming $L =L_{2-10~{\rm keV}}$ during the Flare, 
and $R ={3\times 10^{13}}$), we estimated a value of $l\sim$10 during the Flare period 
(corresponding to a pair optical depth $\simeq$0.1, Stern et al. 1995), 
indicating that, at least during the most compact flares, the pair contribution may not be entirely 
negligible for the source spectral evolution (Haardt et al. 1997).

If such alternative mechanisms are indeed responsible for the time lag observed during the Flare, 
the Comptonization timescale must be correspondingly shorter. In this scenario, 
the estimate of the electron density of the Flare region given in \S6.1 
should be considered a lower limit.

\

Finally, we anticipate that in the future more precise timing analysis will be able 
to discriminate among these alternative scenarios. In fact, models in which 
(short- and long-term) spectral variations are driven by only Comptonization and/or 
disk-corona reverberation effects should exhibit only soft-to-hard time lags, contrary 
to other models that may have zero lag and/or hard-to-soft time lags. 

\section{Conclusions} 

The model-independent timing analysis of a $\sim$ 95 ks 
long $XMM-Newton$ observation of MCG-6-30-15 presented in this work has shown that:

(1) The fractional variability around the FeK line energy is ``resolved'' with unprecedented 
detail into two different components: a narrow component of lower variability at 
E $\sim$6.4 keV consistent with its being produced by the outer parts of the accretion disk 
and a broad component in the 4.7-5.8 keV energy band, significantly more variable than the 
continuum itself (Ballantyne et al. 2003).
Whether this second component is due to emission or absorption is unclear, but 
it is likely associated to a highly variable and redshifted iron line that probes  
the inner regions of the accretion disk.

(2) The reduced variability of the source continuum at soft (E $\lsimeq$600 eV) energies 
is consistent with a gradual onset of soft thermal emission from the 
outer parts of the accretion disk.
The reduced variability at higher (E $\gsimeq$2 keV) energies is possibly due to a  
power law having a pivot point at E $\gsimeq$10 keV, or to a gradually increasing 
contribution from a reflection component.

(3) During a remarkably strong Flare, a significant soft-to-hard 
time lag of $\sim$600 s is detected. It is the first time such a time lag has been detected in 
MCG-6-30-15 on such a short timescale. We find it to be consistent with a delay due 
to Comptonization up-scattering. On this hypothesis, we are able to infer 
estimates of the electron density and dimensions of the Flare region:
$\rho_{e^-} \simeq 2.5-25\times 10^{10}$ cm$^{-3}$ and $R_{{\rm Flare}} \sim 6\times 10^{12}$cm 
$\sim 40~ r_{\rm g}$, respectively.

(4) Ratios of spectra taken during and after the Flare show evidence that the FeK line 
intensity has varied very significantly, and may have 
done so in response to the Flare with a $\sim$1500-3000 s lag. 
In the simplest scenario, in which the Flare is produced in the corona 
and the FeK line is reflected from the disk, the observed lag allows us 
to estimate the disk-corona distance as $\lsimeq$$4\times 10^{13}$ cm $\sim 240~ r_{\rm g}$.

(5) The present short-term-variability results well illustrate the potential offered by a high-throughput 
instrument like $XMM-Newton$ for obtaining constraints on theoretical emission models, 
and map the inner regions of AGNs.

\begin{acknowledgements}
We thank S. Molendi, N. La Palombara, G. Giovannini, P. Grandi, A. Malizia, L. Foschini, G. G. C. Palumbo, 
G. Ghisellini, F. Haardt and P. O. Petrucci for useful conversations.
This work has been partly financed by ASI under contracts: I/R/042/02 and I/R/045/02.
\end{acknowledgements}

\onecolumn
\appendix
\section{Definition of the RMS spectrum}
\label{sec:appendix:estimators}

We used an innovative fractional variability function, called ``RMS spectrum''. It allows 
a reduction of the associated errors, and therefore a finer energy binning, when a sufficiently 
high number of counts is available.
If on this condition of high statistics per bin, one can consider negligible 
the Poisson noise, the standard fractional variability amplitude 
(F$_{\rm var}$, e.g. Nandra et al. 1997, Edelson et al. 2002), can be rewritten as:
\begin{equation}
RMS(E)= \frac{\displaystyle \sqrt{\sum_{i=1}^{N}\frac{[x(E,\Delta t_i)-<x(E)>]^2}{N-1}}}{<x(E)>},
\end{equation}
where $x(E,\Delta t_i)$ is the source count rate in the energy band $E$ 
during the time interval $\Delta t_i$, $<x(E)>$ is the mean source count rate at energy $E$ 
during the whole observation, N is the number of time bins (i.e., N$= \frac{Exposure Time}{\Delta t_i}$).
The RMS spectrum characterizes the source fractional variability as a function of the energy. 
Compared to the more widely used F$_{\rm var}$, the RMS(E) has the advantage that its associated errors 
can be calculated analytically by the standard propagation equation, yielding 
\begin{equation}
\displaystyle
\sigma_{\rm RMS} = {\bfseries \sqrt{\sum^N_{i=1}\biggl[\frac{\partial RMS}{\partial x(E,\Delta t_i)} dx(E,\Delta t_i) \biggr]^2 } =} \frac{\sqrt{\displaystyle \sum_{i=1}^N[(x(E,\Delta t_i)-<x(E)>)\sigma_{x(E,\Delta t_i)}]^2}}
{\displaystyle <x(E)>\sqrt{(N-1) \sum^N_{i=1}[x(E,\Delta t_i)-<x(E)>]^2}}.
\end{equation}
where $\sigma_{x(E,\Delta t_i)}$ is the error associated to $x(E,\Delta t_i)$, and where we assumed that 
the error associated to $<x(E)>$ is negligible compared to $\sigma_{x(E,\Delta t_i)}$.

This assessment of the error is less ``conservative'' than using the more widely used fractional 
variability equation (Fig. A.1, left panel) since it requires that the energy and/or 
the time bins used in the calculation be chosen so as to have a negligible random Poisson noise.
Under these conditions, the error given in Eq. A.2 (Fig. A.1, right panel) 
is also very similar to the one calculated recently by Vaughan et al. (2003b) using Monte Carlo 
simulations (see Fig. A.1, middle panel and see Eq. B.3 in Vaughan et al. 2003b).
Its greatest advantage is however that in cases like the present one of large signal-to-noise, it does 
reduce significantly the errors and thus allows a much finer sampling of the energy and/or time scales
involved (cf. Edelson et al. 2002 for the overestimation of the errors caused by a
conservative approach in the calculation of the standard fractional variability function).
Furthermore, since the errors and the RMS data points are formaly independent 
(unlike those calculated by Vaughan et al. 2003b) we could estimate 
the significance of the variability structures fitting directly the RMS spectrum (see \S 3.1).
In particular, in the present data, we chose energy and time bins in order to have at least 
350 counts per bin, so as to exclude any Poisson noise (which we found to be relevant only for 
$<$200 counts per bin).
\begin{figure*} [!h]
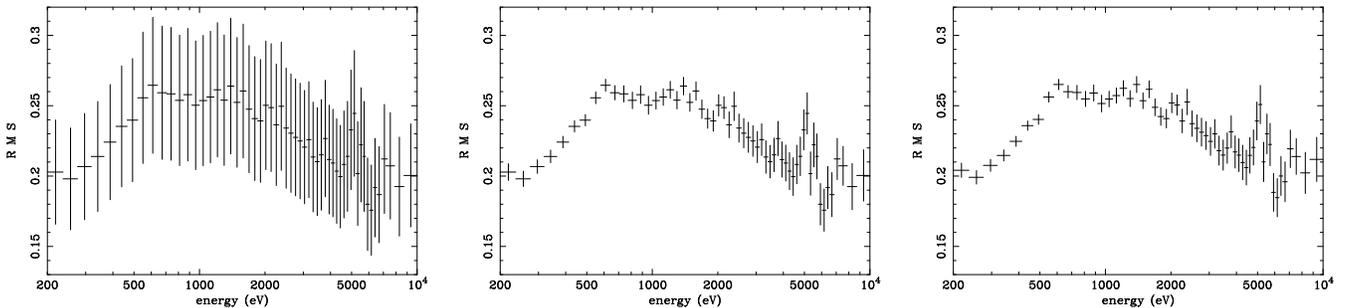

\begin{tabular}{c c c}
\psfig{file=sppn06146ex000.ps,width=5.6cm,height=4cm,angle=-90} & \psfig{file=sppn06146fv000.ps,width=5.6cm,height=4cm,angle=-90} & \psfig{file=sppn06146sm100.ps,width=5.6cm,height=4cm,angle=-90} \\
\end{tabular}
\caption{RMS spectra calculated with the equations used in Edelson et al. 2002 (Left panel), 
in Vaughan et al. 2003b (Middle panel) and in this work (Right panel).
The main difference of the RMS functions is their error estimates, as follows: Edelson et al. (2002) used 
a conservative approach in the calculation of the standard fractional variability function, leading to 
an overestimation of the errors; Vaughan et al. (2003b) estimate the errors with Monte-Carlo simulations; 
the formula presented here is a good approximation of the Vaughan et al. results if Poisson noise 
is negligible (i.e. if $>$ 300 counts/bin). 
The advantage of the equation presented by Vaughan et al. is that it is always valid, 
also in the case of few counts/bin, but its disadvantage is that it cannot be formally be fitted to estimate 
the significance of structures, since its errors are not independent.
The advantage of the RMS equation presented in this work is that the errors are independent, 
so fine structures can be fitted. 
The disadvantage of this method is that it is applicable only if the statistics are good enough to ignore Poisson noise
($\gsimeq$ 300 counts/bins).}
\end{figure*}
Finally, we checked that the results obtained using the RMS spectrum are entirely consistent with 
those obtained using a standard fractional variability formula. But thanks to the lower error bars, we could 
detect the fine structures with increased significance (see text).

{}

%\bibitem{} Arnaud K.A., 1996, Astronomical Data Analysis Software and Systems 
%V, eds. Jacoby G. and Barnes J., p17, ASP Conf. Series volume 101

%\vfill\eject
%\onecolumn


\begin{thebibliography}{}
\bibitem[Ballantyne, Vaughan, \& Fabian(2003)]{2003MNRAS.342..239B} Ballantyne, D.~R., Vaughan, S., \& Fabian, A.~C.\ 2003, \mnras, 342, 239 
\bibitem[Beloborodov(1999)]{1999ApJ...510L.123B} Beloborodov, A.~M.\ 1999a, \apjl, 510, L123 
\bibitem[Beloborodov(1999)]{1999hepa.conf..295B} Beloborodov, A.~M.\ 1999b, ASP Conf.~Ser.~161: High Energy Processes in Accreting Black Holes, 295 (astro-ph/9901108)
\bibitem[Branduardi-Raymont et al.(2001)]{2001A&A...365L.140B} Branduardi-Raymont, G., Sako, M., Kahn, S.~M., Brinkman, A.~C., Kaastra, J.~S., \& Page, M.~J.\ 2001, \aap, 365, L140 
\bibitem[di Matteo, Celotti, \& Fabian(1999)]{1999MNRAS.304..809D} di Matteo, T., Celotti, A., \& Fabian, A.~C.\ 1999, \mnras, 304, 809 
\bibitem[Edelson \& Krolik(1988)]{1988ApJ...333..646E} Edelson, R.~A.~\& Krolik, J.~H.\ 1988, \apj, 333, 646 
\bibitem[Edelson et al.(2001)]{2001ApJ...554..274E} Edelson, R., Griffiths, G., Markowitz, A., Sembay, S., Turner, M.~J.~L., \& Warwick, R.\ 2001, \apj, 554, 274 
\bibitem[Edelson et al.(2002)]{2002ApJ...568..610E} Edelson, R., Turner, T.~J., Pounds, K., Vaughan, S., Markowitz, A., Marshall, H., Dobbie, P., \& Warwick, R.\ 2002, \apj, 568, 610 
\bibitem[Fabian, Rees, Stella, \& White(1989)]{1989MNRAS.238..729F} Fabian, A.~C., Rees, M.~J., Stella, L., \& White, N.~E.\ 1989, \mnras, 238, 729 
\bibitem[Fabian, Iwasawa, Reynolds, \& Young(2000)]{2000PASP..112.1145F} Fabian, A.~C., Iwasawa, K., Reynolds, C.~S., \& Young, A.~J.\ 2000, \pasp, 112, 1145 
\bibitem[Fabian et al.(2002)]{2002MNRAS.335L...1F} Fabian, A.~C., Vaughan, S., Nandra, K.~et al.\ 2002, \mnras, 335, L1 
\bibitem[Fabian \& Vaughan(2003)]{2003MNRAS.340L..28F} Fabian, A.~C.~\& Vaughan, S.\ 2003, \mnras, 340, L28 
\bibitem[Ghisellini, Haardt, \& Fabian(1993)]{1993MNRAS.263L...9G} Ghisellini, G., Haardt, F., \& Fabian, A.~C.\ 1993, \mnras, 263, L9 
\bibitem[Guainazzi et al.(1999)]{1999A&A...341L..27G} Guainazzi, M., Matt, G., Molendi, S.~et al.\ 1999, \aap, 341, L27 
\bibitem[Haardt \& Maraschi(1991)]{1991ApJ...380L..51H} Haardt, F.~\& Maraschi, L.\ 1991, \apjl, 380, L51 
\bibitem[Haardt \& Maraschi(1993)]{1993ApJ...413..507H} Haardt, F.~\& Maraschi, L.\ 1993, \apj, 413, 507 
\bibitem[Haardt, Maraschi, \& Ghisellini(1994)]{1994ApJ...432L..95H} Haardt, F., Maraschi, L., \& Ghisellini, G.\ 1994, \apjl, 432, L95 
\bibitem[Haardt(1997)]{1997MmSAI..68...73H} Haardt, F.\ 1997, Memorie della Societa Astronomica Italiana, 68, 73 
\bibitem[Haardt, Maraschi, \& Ghisellini(1997)]{1997ApJ...476..620H} Haardt, F., Maraschi, L., \& Ghisellini, G.\ 1997, \apj, 476, 620 
\bibitem[Iwasawa et al.(1996)]{1996MNRAS.282.1038I} Iwasawa, K., Fabian, A. C., Reynolds, C. S.~et al.\ 1996, \mnras, 282, 1038 
\bibitem[Jansen et al.(2001)]{2001A&A...365L...1J} Jansen, F., Lumb, D., Altieri, B.~et al.\ 2001, \aap, 365, L1 
\bibitem[La Palombara(2002)]{2002AGN05.conf} La Palombara, N., Molendi, S., Wilms J., Reynolds, C.S., 2002, AGN05 Conf.: Inflows, Outflows and Reprocessing around black holes, (astro-ph/0210357)
\bibitem[Lee et al.(2000)]{2000MNRAS.318..857L} Lee, J.~C., Fabian, A.~C., Reynolds, C.~S., Brandt, W.~N., \& Iwasawa, K.\ 2000, \mnras, 318, 857 
\bibitem[Lee et al.(2001)]{2001ApJ...554L..13L} Lee, J.~C., Ogle, P.~M., Canizares, C.~R., Marshall, H.~L., Schulz, N.~S., Morales, R., Fabian, A.~C., \& Iwasawa, K.\ 2001, \apjl, 554, L13 
\bibitem[Lee et al.(2002)]{2002ApJ...570L..47L} Lee, J.~C., Iwasawa, K., Houck, J.~C., Fabian, A.~C., Marshall, H.~L., \& Canizares, C.~R.\ 2002, \apjl, 570, L47 
\bibitem[Longinotti et al.(2003)]{2003A&A...410..471L} Longinotti, A.~L., Cappi, M., Nandra, K., Dadina, M., \& Pellegrini, S.\ 2003, \aap, 410, 471 
\bibitem[Miniutti et al.(2003)]{2003MNRAS.338..389M} Miniutti, G., Fabian, A.~C., Goyder, R., \& Lasenby, A.~N.,\ 2003, \mnras, in press 
\bibitem[Mushotzky, Done, \& Pounds(1993)]{1993ARA&A..31..717M} Mushotzky, R.~F., Done, C., \& Pounds, K.~A.\ 1993, \araa, 31, 717 
\bibitem[Nandra \& Pounds(1994)]{1994MNRAS.268..405N} Nandra, K.~\& Pounds, K.~A.\ 1994, \mnras, 268, 405 
\bibitem[Nandra et al.(1997)]{1997ApJ...476...70N} Nandra, K., George, I.~M., Mushotzky, R.~F., Turner, T.~J., \& Yaqoob, T.\ 1997, \apj, 476, 70 
\bibitem[Nandra et al.(1999)]{1999ApJ...523L..17N} Nandra, K., George, I.~M., Mushotzky, R.~F., Turner, T.~J., \& Yaqoob, T.\ 1999, \apjl, 523, L17 
\bibitem[Negoro, Matsuoka, \& Mihara(1999)]{1999AN....320..313N} Negoro, H., Matsuoka, M., \& Mihara, T.\ 1999, Astronomische Nachrichten, 320, 313 
\bibitem[Negoro et al.(2000)]{2000AdSpR..25..481N} Negoro, H., Matsuoka, M., Mihara, T., Otani, C., Wang, T.~G., \& Awaki, H.\ 2000, Advances in Space Research, 25, 481 
\bibitem[Otani et al.(1996)]{1996PASJ...48..211O} Otani, C., Kii, T., Reynolds, C. S.~et al.\ 1996, \pasj, 48, 211 
\bibitem[Perola et al.(2002)]{2002A&A...389..802P} Perola, G.~C., Matt, G., Cappi, M., Fiore, F., Guainazzi, M., Maraschi, L., Petrucci, P.~O., \& Piro, L.\ 2002, \aap, 389, 802 
\bibitem[Petrucci \& Dadina(2003)]{2003} Petrucci, P.~O.~\& Dadina, M.\ 2003 in preparation
\bibitem[Pounds \& Reeves(2002)]{2002.conf} Pounds, K., Reeves, J., 2002, ESTEC Conf.: New Visions of the X-ray Universe in the XMM-Newton and Chandra Era, (astro-ph/0201436)
\bibitem[Poutanen \& Svensson(1996)]{1996ApJ...470..249P} Poutanen, J.~\& Svensson, R.\ 1996, \apj, 470, 249 
\bibitem[Poutanen \& Fabian(1999)]{1999MNRAS.306L..31P} Poutanen, J.~\& Fabian, A.~C.\ 1999, \mnras, 306, L31 
\bibitem[Poutanen(2001)]{2001AdSpR..28..267P} Poutanen, J.\ 2001, Advances in Space Research, 28, 267 (astro-ph/0102325)
\bibitem[Press, Teukolsky, Vetterling, \& Flannery(1992)]{1992nrca.book.....P} Press, W.~H., Teukolsky, S.~A., Vetterling, W.~T., \& Flannery, B.~P.\ 1992, Cambridge: University Press, Numerical Recipes in C++ 1992, 2nd ed.,
\bibitem[Ruszkowski \& Fabian(2000)]{2000MNRAS.315..223R} Ruszkowski, M.~\& Fabian, A.~C.\ 2000, \mnras, 315, 223 
\bibitem[Sako et al.(2003)]{2003ApJ...596..114S} Sako, M., Kahn, S. M., Branduardi-Raymont, G.~et al.\ 2003, 
\apj, 596, 114 
\bibitem[Shih, Iwasawa, \& Fabian(2002)]{2002MNRAS.333..687S} Shih, D.~C., Iwasawa, K., \& Fabian, A.~C.\ 2002, \mnras, 333, 687 
\bibitem[Stern et al.(1995)]{1995ApJ...449L..13S} Stern, B.~E., Poutanen, J., Svensson, R., Sikora, M., \& Begelman, M.~C.\ 1995, \apjl, 449, L13 
\bibitem[Str{\" u}der et al.(2001)]{2001A&A...365L..18S} Str{\" u}der, L. Briel, U., Dennerl, K.~et al.\ 2001, \aap, 365, L18 
\bibitem[Sunyaev \& Titarchuk(1980)]{1980A&A....86..121S} Sunyaev, R.~A.~\& Titarchuk, L.~G.\ 1980, \aap, 86, 121 
\bibitem[Turner et al. 2003]{1980A&A....86..121S} Turner, A.~K., Fabian, A.~C., Vaughan, S., \& Lee, J.~C.\ 2003, \mnras, in press 
\bibitem[Vaughan, Fabian, \& Nandra(2003)]{2003MNRAS.339.1237V} Vaughan, S., Fabian, A.~C., \& Nandra, K.\ 2003a, \mnras, 339, 1237 
\bibitem[Vaughan, Edelson, Warwick, \& Uttley(2003)]{2003MNRAS.339.1237V} Vaughan, S., Edelson, R., Warwick, R.~S.~\& Uttley, P.\ 2003b, \mnras, in press 
\bibitem[White \& Peterson(1994)]{1994PASP..106..879W} White, R.~J.~\& Peterson, B.~M.\ 1994, \pasp, 106, 879 
\bibitem[Wilms et al.(2001)]{2001MNRAS.328L..27W} Wilms, J., Reynolds, C.~S., Begelman, M.~C., Reeves, J., Molendi, S., Staubert, R., \& Kendziorra, E.\ 2001, \mnras, 328, L27 
\end{thebibliography}
\end{document}